\documentclass[]{JHEP3}
\usepackage{epsfig,multicol}
\usepackage{graphicx}
\usepackage{xspace}
\newcommand{\sla}{\kern -5.4pt /}
\newcommand{\Dir}{\kern -6.4pt\Big{/}}
\newcommand{\Dirin}{\kern -10.4pt\Big{/}\kern 4.4pt}
\newcommand{\DDir}{\kern -7.6pt\Big{/}}
\newcommand{\DGir}{\kern -6.0pt\Big{/}}

\newcommand{\be}{\begin{equation}}
\newcommand{\ee}{\end{equation}}
\newcommand{\bea}{\begin{eqnarray}}
\newcommand{\eea}{\end{eqnarray}}
\newcommand{\beanon}{\begin{eqnarray*}}
\newcommand{\eeanon}{\end{eqnarray*}}
\newcommand{\ba}{\begin{array}}
\newcommand{\ea}{\end{array}}
\newcommand{\bd}{\begin{description}}
\newcommand{\ed}{\end{description}}
\newcommand{\bi}{\begin{itemize}}
\newcommand{\ei}{\end{itemize}}
\newcommand{\ben}{\begin{enumerate}}
\newcommand{\een}{\end{enumerate}}
\newcommand{\bc}{\begin{center}}
\newcommand{\ec}{\end{center}}

\newcommand{\GeV}{\mbox{${\mathrm GeV}$}\xspace}

\newcommand{\slaRainwater}[1]{/\!\!\!#1}

\newcommand{\ordEW}{$\mathcal{O}(\alpha_{\scriptscriptstyle EM}^6)$\xspace}
\newcommand{\ordQCD}{$\mathcal{O}(\alpha_{\scriptscriptstyle EM}^4
  \alpha_{\scriptscriptstyle S}^2)$\xspace}
\newcommand{\ordQCDsq}{$\mathcal{O}(\alpha_{\scriptscriptstyle EM}^2
  \alpha_{\scriptscriptstyle S}^4)$\xspace}

\newcommand{\eqn}[1]{Eq.(\ref{#1})}
\newcommand{\eqns}[2]{Eqs.(\ref{#1}--\ref{#2})}

\newcommand{\fig}[1]{Fig.~\ref{#1}}

\newcommand{\sect}[1]{Sect.~\ref{#1}}

\newcommand{\diags}[2]{diagrams~({#1})--({#2})}

\newcommand{\Phantom}{{\tt PHANTOM}\xspace}

\newcommand{\MadEvent}{{\tt MADEVENT}\xspace}

\def\pl #1 #2 #3 {{\it Phys.~Lett.} {\bf#1} (#2) #3}   
\def\np #1 #2 #3 {{\it Nucl.~Phys.} {\bf#1} (#2) #3}
\def\zp #1 #2 #3 {{\it Z.~Phys.} {\bf#1} (#2) #3}
\def\pr #1 #2 #3 {{\it Phys.~Rev.} {\bf#1} (#2) #3}
\def\prep #1 #2 #3 {{\it Phys.~Rep.} {\bf#1} (#2) #3}
\def\prl #1 #2 #3 {{\it Phys.~Rev.~Lett.} {\bf#1} (#2) #3}
\def\intj #1 #2 #3 {{\it Int. J. Mod. Phys.} {\bf#1} (#2) #3}
\def\mpl #1 #2 #3 {{\it Mod.~Phys.~Lett.} {\bf#1} (#2) #3}
\def\rmp #1 #2 #3 {{\it Rev. Mod. Phys.} {\bf#1} (#2) #3}
\def\cpc #1 #2 #3 {{\it Comp. Phys. Commun.} {\bf#1} (#2) #3}
\def\epj #1 #2 #3 {{\it Eur. Phys. J.} {\bf#1} (#2) #3}
\def\jhep #1 #2 #3 {{\it JHEP} {\bf#1} (#2) #3}

\title{A new analysis of $PP \rightarrow b\bar{b}\ell \nu jj$ at the LHC:
Higgs and $W$ boson associated production with two tag jets.}

\author{
Alessandro Ballestrero$^a$,
Giuseppe Bevilacqua$^{a,b}$ and
Ezio Maina$^{a,b}$\\
$^a$ INFN, Sezione di Torino, Italy,\\
$^b$ Dipartimento di Fisica Teorica, Universit\`a di Torino, Italy
}

\preprint{DFTT 17/2008}

\abstract{
Higgs production in association with a $W$ boson and two energetic tag jets at
the LHC is studied for $M_H = 120$ \GeV, with the Higgs decaying to
$b\bar{b}$ and the $W$ to $\ell\nu$ ($\ell = e,\,\mu$),
guaranteeing high trigger efficiency.
All parton level backgrounds are analyzed,
including the effect of fake $b$--tagging.
We discuss two detection strategies: in the first, more traditional,
one, two jets are required to be
$b$--tagged while in the second, which has not been previously examined
in detail, only one tag is required. After all
selection cuts about 80 and 200
events are foreseen  in the two cases for a standard luminosity
of 100 $fb^{-1}$
with a S/B ratio of 1/25 and 1/60 respectively. The corresponding statistical
significancies, S/$\sqrt{\mathrm B}$ are 1.81 and 1.82.
}


\begin{document}

\section{Introduction}
\label{sec:intro}

Higgs couplings to fermions are predicted unambigously in all chiral
theories in
which Electroweak Symmetry Breaking is realized through the Higgs mechanism.
A fundamental consequence of this general scheme is the proportionality of the
Yukawa couplings to the corresponding fermion masses.
A check of this proportionality is a fundamental test of EWSB.

The coefficient of proportionality between the Yukawa couplings and the
fermion masses depend on the structure of the Higgs sector. For instance they
are different in the Standard Model and in the MSSM. Therefore a measurement of
these couplings would be extremely useful in distinguishing among competing
theories

The $H\rightarrow b\bar{b}$ channel is of crucial relevance. It is by far the largest
Higgs boson decay channel for a light Higgs, $M_H < 140$ \GeV.
It suffers from huge QCD
backgrounds and it is necessary to consider production mechanisms which
enhance Higgs production in comparison with QCD $b\bar{b}$ production in order
to have a fighting chance to its detection.
This typically involves additional particles or energetic jets in the final
state.  

A detailed discussion of the extraction of the Higgs couplings to gauge
bosons and fermions and extensive references to previous literature can be found
in \cite{ZKNR,Duhrssen:2004cv}.

A number of channels have been studied in the past:

\bi
\item  $t\bar{t} b\bar{b}$ \cite{ATLAS,ttH1,ttH2,ttH3}: this channel provides
spectacular events with four $b$'s, one high $p_T$ lepton and two additional
jets with a cross section of about 200 $fb$ for $M_H = 200$ \GeV. It suffers
however from a large combinatorial background due to the very presence of four
$b$'s in the final state which makes it difficult to reconstruct the Higgs peak.
Earlier analysis \cite{ATLAS,ttH1,ttH2} were rather optimistic claiming
a statistical
significance of about $3\div4$ for an integrated luminosity of 30 $fb^{-1}$.
A more recent analysis \cite{ttH3}, using full detector simulation, is more
cautious, obtaining a statistical significance of order 1.5 for a 
luminosity of 60 $fb^{-1}$, which should strike a cautionary note concerning
the conclusions of all papers, including the present one, which do not
incorporate the full reconstruction chain.

\item $b\bar{b}Wjj $\cite{Rainwater00}: this channel was studied in a  
Vector Boson Fusion like regime, characterized by large separation and large
mass of the two tag jets, and compared to the background due to
$W b\bar{b}jj$ non resonant QCD production, and $t\bar{t} + 2 jets$ events,
where both $W$'s from the 
top quarks decay leptonically ($e$ or $\mu$) and one of the leptons is too low 
in $p_T$ to be observed. The presence of an isolated, high $p_T$ lepton
guarantees high triggering efficiency.
Very interesting signal to background ratios were reported.
 
\item $b\bar{b}jj$ \cite{Mangano:2002wn}: Vector Boson Fusion like cuts allow
the extraction of the signal, but the required luminosity is in the range of
about $600 fb^{-1}$.

\item $b\bar{b}\gamma jj$ \cite{bbjjgamma}: the requirement of an additiona
high--$p_T$ photon increases substantially the signal to background ratio.
This channel was analyzed and compared with the one in \cite{Mangano:2002wn}
using the same selection criteria, obtaining significancies in the range between
one and three for a standard high luminosity of 100 $fb^{-1}$ and Higgs masses
between 120 and 140 \GeV .

\ei
 
In the following we reanalyze the reaction 
$PP \rightarrow b\bar{b}W(\ell\nu)jj$ as a mean to detect the Higgs decay to
$b\bar{b}$ for $M_H = 120$ \GeV .
In the next Section we discuss the analysis performed in \cite{Rainwater00}
and propose a number of possible
improvements, while in \sect{sec:calc} the main features
of the calculation are shown. Then we present our results: in \sect{sec:2b}
employing the traditional approach which requires both $b$'s from the Higgs decay
to be tagged and in \sect{sec:1b} requiring only a single $b$--tag.
Finally we summarize the main points of our discussion.

\section{$PP \rightarrow H(b\bar{b})W(\ell\nu)jj$ and its backgrounds}
\label{sec:discuss}

For convenience we recall here the main features of the study performed in
\cite{Rainwater00}.
The $WHjj$ events considered were pure EW processes. The QCD contribution to
$WHjj$ was conservatively neglected.
The main background processes were taken to be nonresonant QCD 
$W b\bar{b}jj$ production, and $t\bar{t}+ 2 jets$ events, where both $W$'s from the 
top quarks decay leptonically ($e$ or $\mu$) and one of the leptons is soft and undetected,
the limit being set at $p_T(l,min) < 10$~GeV.

The following set of cuts were used:

\bea
\label{eq:gap}
& p_{T_j} \geq 30~{\rm GeV} \, , \; \; |\eta_j| \leq 5.0 \, , \; \;
\triangle R_{jj} \geq 0.6 \, , \nonumber \\
& p_{T_b} \geq 15~{\rm GeV} \, , \; \; |\eta_b| \leq 2.5 \, , \; \;
\triangle R_{jb} \geq 0.6 \, , \\
& p_{T_\ell} \geq 20~{\rm GeV} \, ,\; \;
|\eta_{\ell}| \leq 2.5 \, , \; \; 
\triangle R_{j\ell,b\ell} \geq 0.6 \, , \nonumber \\
& \eta_{j,min} + 0.7 < \eta_{b,\ell} < \eta_{j,max} - 0.7 \, , \nonumber \\
& \eta_{j_1} \cdot \eta_{j_2} < 0, \; \; 
\triangle \eta_{tags} = |\eta_{j_1}-\eta_{j_2}| \geq 4.4 \, .\nonumber 
\eea

\be
\label{eq:mjjptb}
M_{j_1 j_2} > 600 \, {\rm GeV} \; , \quad
{p}_T(b_1,b_2) > 50,20 \, {\rm GeV} \; .
\ee

\be
\label{eq:ptmt}
\slaRainwater{p}_T < 100 \, {\rm GeV} \; , \; \; \; 
M_T(\ell,\slaRainwater{p}_T) < 100 \, {\rm GeV} \; .
\ee
where $j = d, u, s, c, g$ while 
$j_1$ and $j_2$ are the tag jets and $b_{1(2)}$ refers
to the $b$--quark with highest (lowest) $p_T$.
For the cuts described in \eqns{eq:gap}{eq:ptmt} and $\ell = e,\, \mu$
the cross sections for $WHjj$,
$Wb\bar{b}jj$, $t\bar{t}jj$ were found to be $1.1$, $4.3$ and $1.2\; fb$
respectively, including the decays of the two bosons, with a signal over
background ratio of 1/5. With an estimated overall efficiency of about $25\%$
and an educated guess concerning the effect of a central mini--jet veto 
$p_{T}^{veto}(j) \geq 20~{\rm GeV}$ of an efficiency of
75\% for the signal and of 30\% for the
background a statistical significance of about 4.4 was foreseen, taking into
account 100 $fb^{-1}$ of data for each of the two experiments.

We propose that a number of issues should receive further attention:

\ben
\item The \ordQCD contribution is potentially large. In fact, 
a simulation of $jj\ell\nu H$, with the Higgs on shell and $M_H = 120$ \GeV,
yields a cross section of 65 fb with the following selection cuts:

\bea
\label{eq:madHW}
& p_{T_j} \geq 20~{\rm GeV} \, , \; \; |\eta_j| \leq 5.0 \, , 
\nonumber \\
& p_{T_b} \geq 20~{\rm GeV} \, , \; \; |\eta_b| \leq 5.0 \, ,  \\
& p_{T_\ell} \geq 20~{\rm GeV} \, ,\; \;
|\eta_{\ell}| \leq 3.0 \, , \nonumber \\
& M_{jj} \geq 60 \, {\rm GeV}  \, , \; \;  M_{bb} \geq 60 \, {\rm GeV} \;.\nonumber 
\eea
With such a relatively large cross section, the impact of this production
channel should be assessed.

\item The analysis has only been performed with cuts optimized for 
$WW$ scattering studies. The Higgs can be produced through boson scattering but
also in Higgs--strahlung which can have quite different kinematic distributions.
It remains to be explored whether other selection procedures are equally or
more effective in extracting a signal.
We are interested in $H \rightarrow b\bar{b}$ regardless of the
details of the production mechanism.

\item The important background due to $t\bar{t}$ production has not been
studied. The semileptonic decay of a
$t\bar{t}$ pair gives exactly the final state we are interested in,
$b\bar{b}\ell \nu jj$, at a prodigious rate compared to the signal.
The top-related background considered in \cite{Rainwater00},
$t\bar{t}jj$ production with one charged lepton lost in the
beampipe turns out to be much smaller than $t\bar{t}$ production even after all
selection cuts, as discussed in \sect{sec:2b}. The
additional jet activity and the particular kinematic signatures expected in top production,
will be effective in reducing both backgrounds
(and should be taken into account for \ordQCD $bbWjj$ production).

\item $b$--tagging is based on several physical observables which discriminate
between jets initiated by $b$'s and jets originating from other kind of partons.
There is however a non negligible probability that $c$--quarks and even
light quarks or gluons produce jets which satisfy the $b$--tagging criteria.
The impact of these fake $b$--taggings has not been taken into account.
Typical values for the
probabilities to pass the $b$--tagging test are:
$\epsilon_b = 0.5$ for a $b$--quark,
$\epsilon_c = 0.1$ for a $c$--quark and $\epsilon_{q/g} = 0.01$ for a light quark or a
gluon, $q = d,u,s$.
In the presence of very large backgrounds as
$t\overline{t}$ and $W+4j$ production, fake hits can have a huge effect
even after severe cuts.

\item The double $b$--tag requirement sharply decreases the expected yield.
The overall detection efficiency is proportional to 
$\epsilon_b^2 = 0.25$ for $\epsilon_b = 0.5$ with double $b$--tagging, while it
becomes $2 \epsilon_b ( 1 - \epsilon_b) + \epsilon_b^2= 0.75$ if at least
one $b$--tagging is
required. In addition the central jet which is supposed to originate from the
Higgs decay, and which is not required to pass the $b$--tagging test,
can be detected in a much larger angular range than the
$b$--tagging coverage, with a further efficiency increase.
It is therefore worthwhile to explore
whether strategies based on single $b$--tagging or no $b$--tagging at all
are viable.   
\een

Therefore, in view of the high relevance of measuring the properties of the
Higgs boson as accurately as possible,
in the following we update and extend the analysis of 
$PP \rightarrow b\bar{b}W(\ell\nu)jj$ including all the features mentioned
above.

\section{Calculation}
\label{sec:calc}

\begin{figure}[htb]
\centering
\mbox{
\includegraphics*[width=0.18\textwidth]{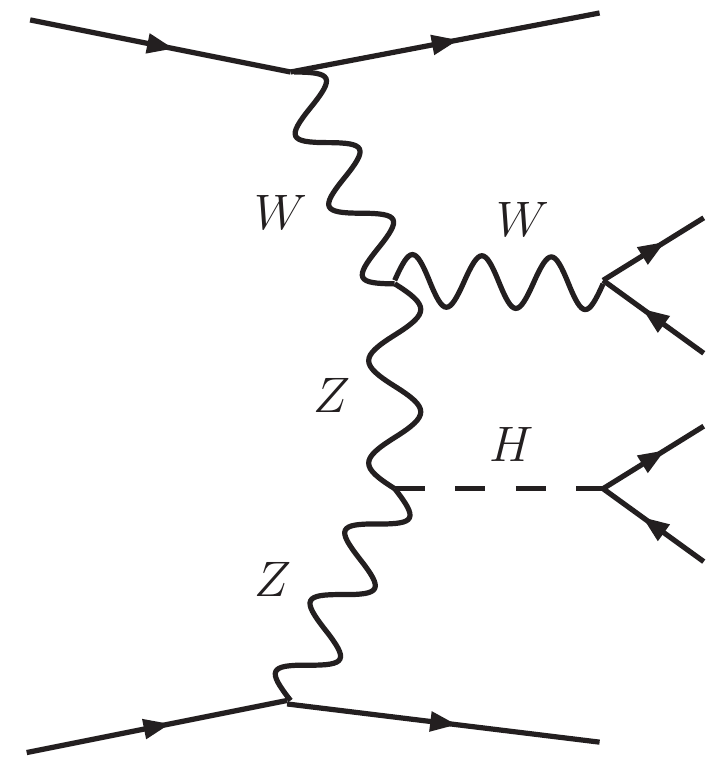}\hspace*{0.5cm}
\includegraphics*[width=0.18\textwidth]{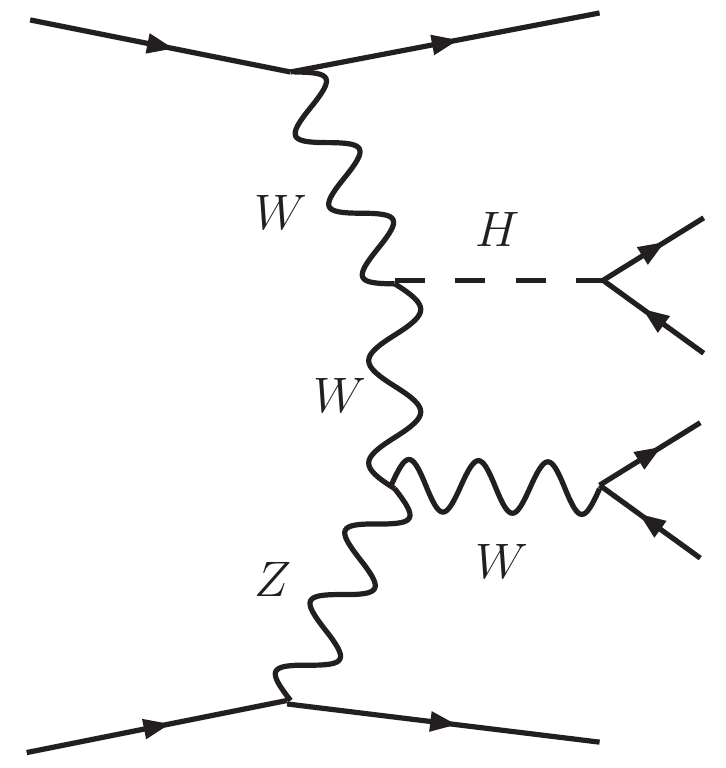}\hspace*{0.6cm}
\includegraphics*[width=0.25\textwidth]{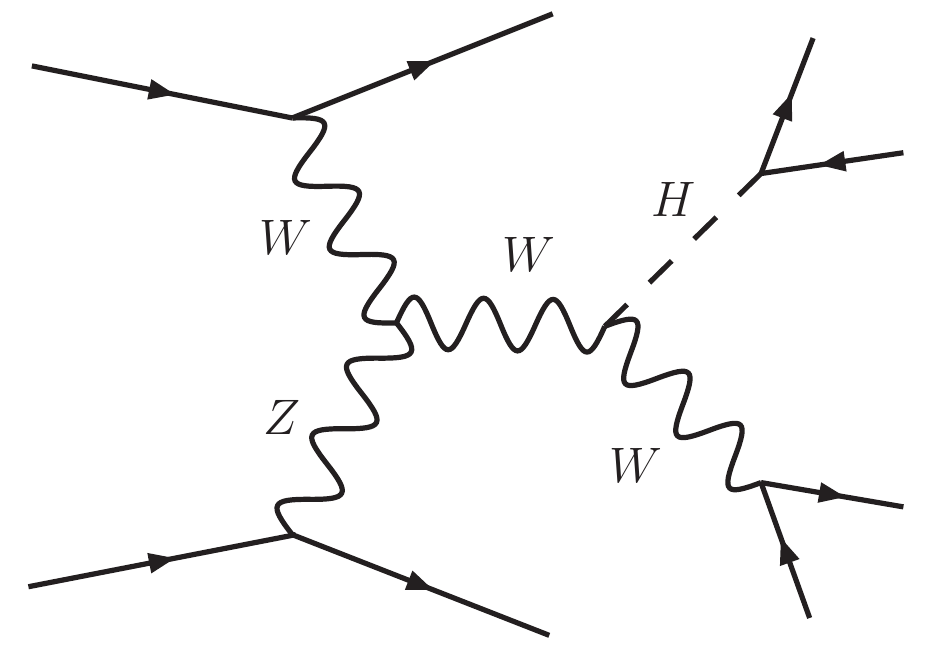}\hspace*{0.6cm}
\includegraphics*[width=0.19\textwidth]{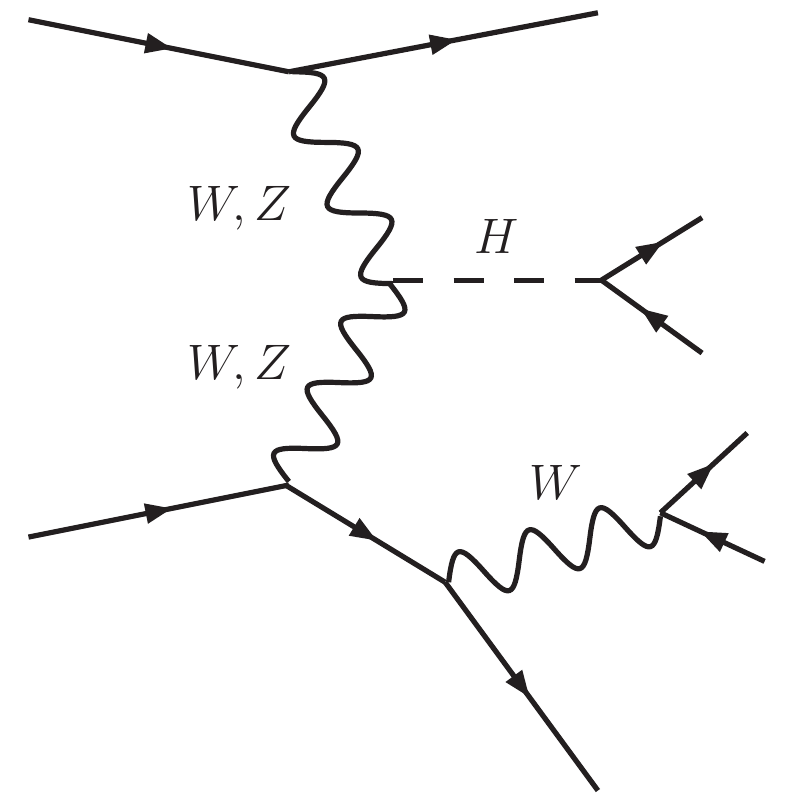}
}\\[1.0cm]
\mbox{
\includegraphics*[width=0.18\textwidth]{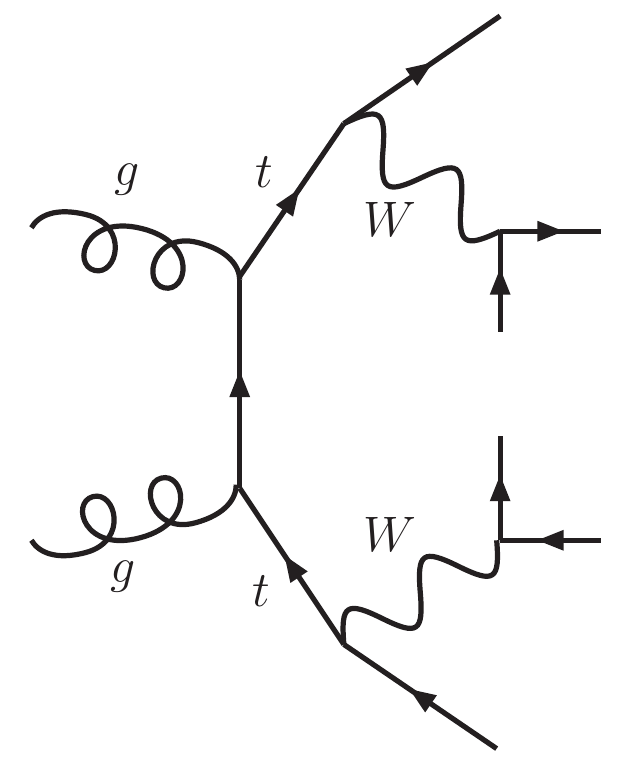}\hspace*{0.5cm}
\includegraphics*[width=0.18\textwidth]{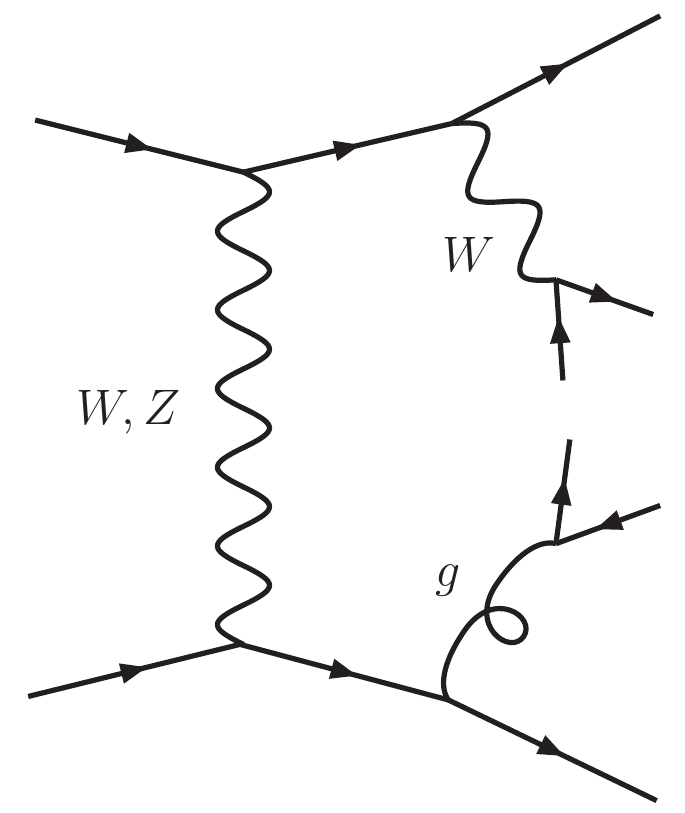}\hspace*{0.6cm}
\includegraphics*[width=0.25\textwidth]{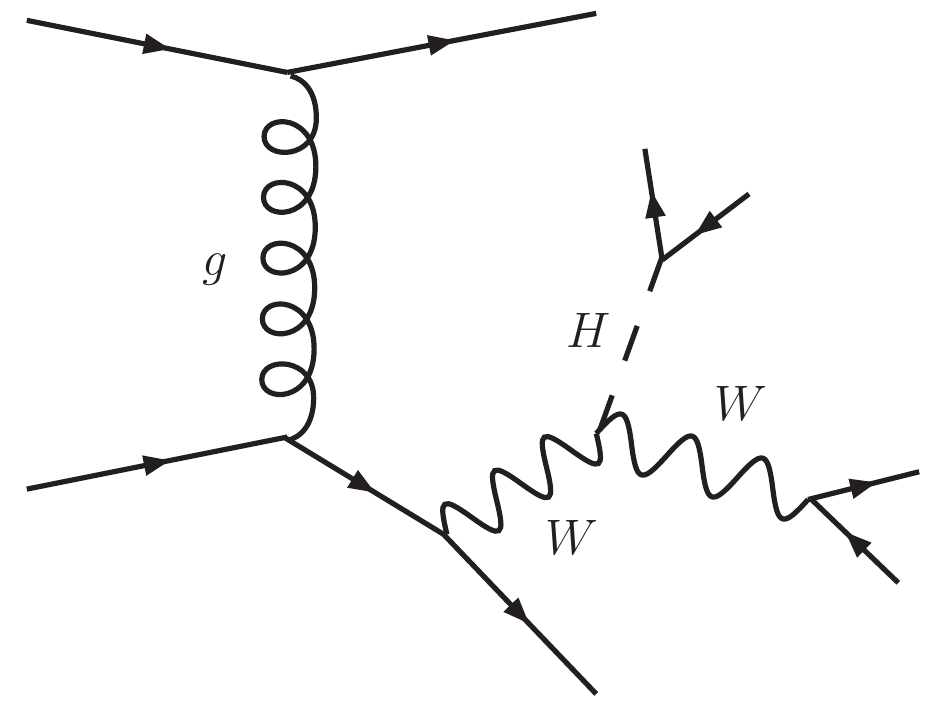}\hspace*{0.6cm}
\includegraphics*[width=0.25\textwidth]{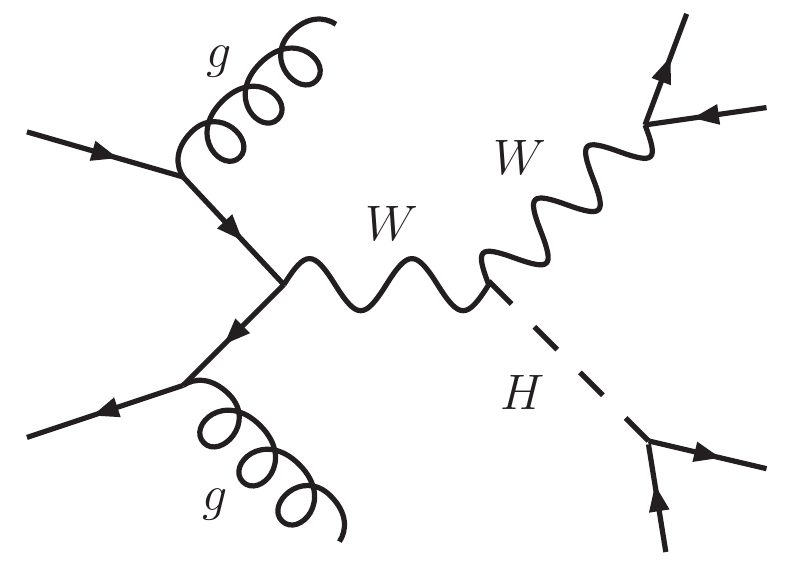}
}\\[1.0cm]
\mbox{
\includegraphics*[width=0.29\textwidth]{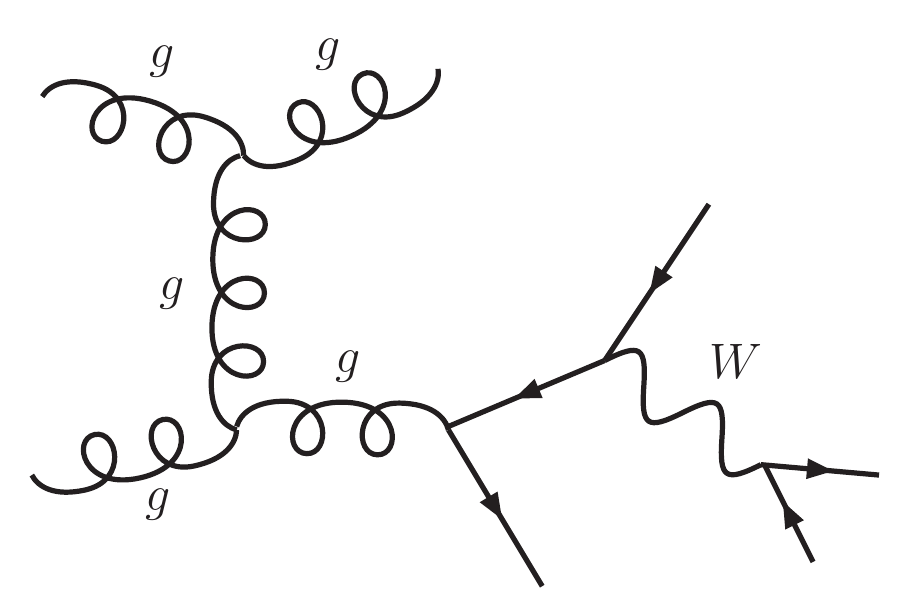}\hspace*{0.5cm}
\includegraphics*[width=0.20\textwidth]{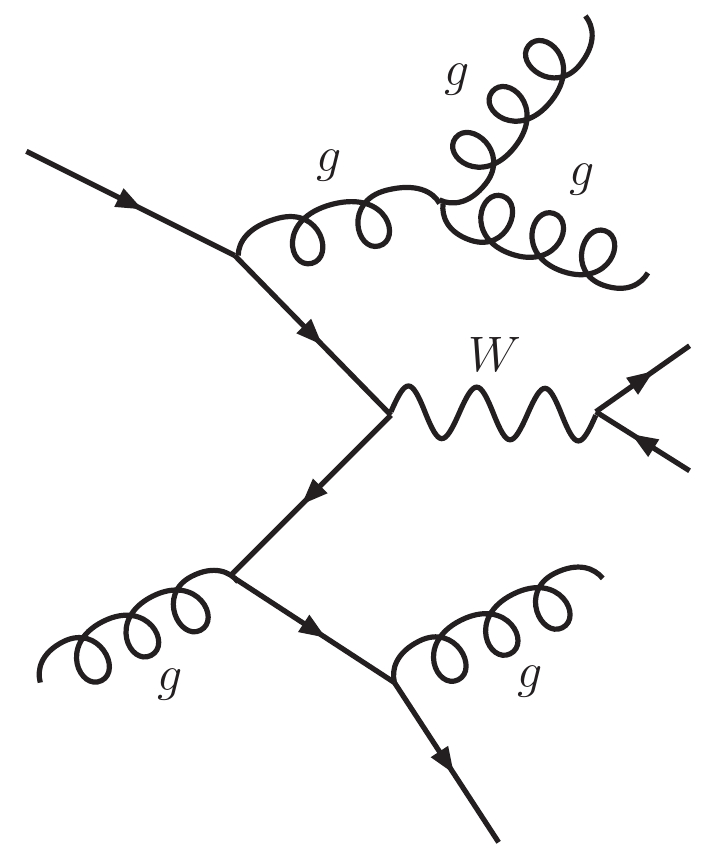}\hspace*{0.6cm}
\includegraphics*[width=0.18\textwidth]{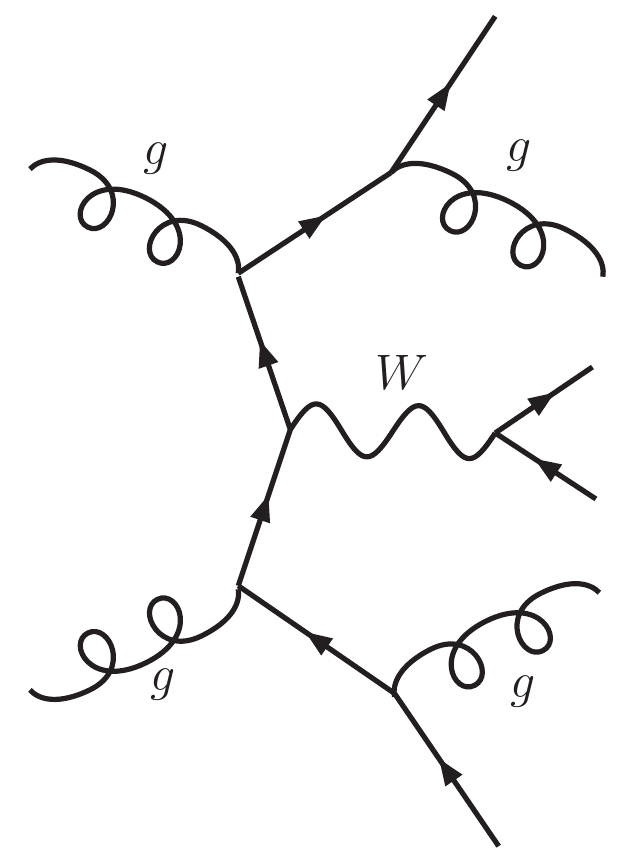}\hspace*{0.6cm}
\includegraphics*[width=0.19\textwidth]{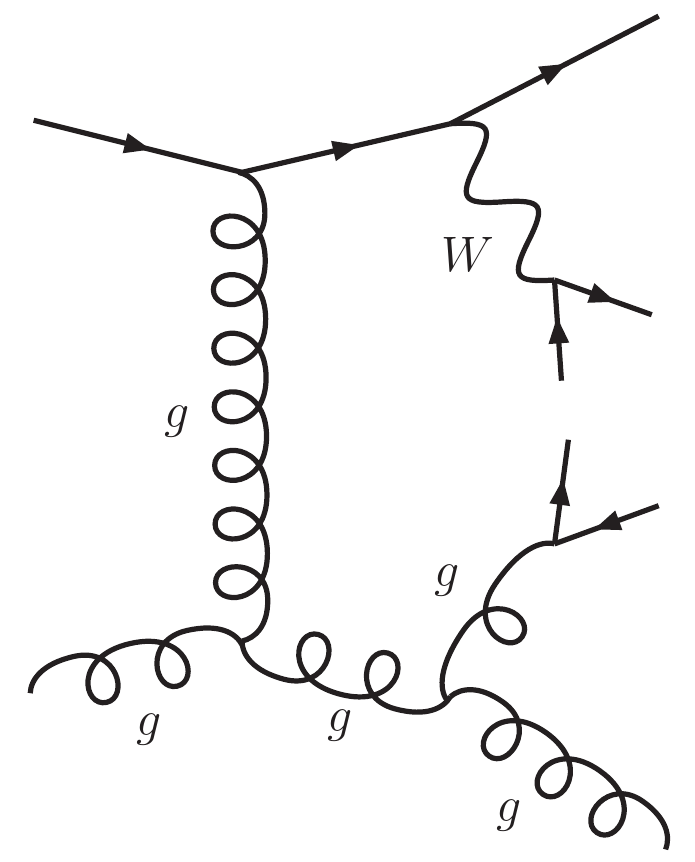}
}\\[0.2cm]
\begin{picture}(0,0) (0,0)
  \put(-180,260) {\small{(a)}}
  \put(-80,260) {\small{(b)}}
  \put(20,260) {\small{(c)}}
  \put(150,260) {\small{(d)}}
  \put(-180,130) {\small{(e)}}
  \put(-90,130) {\small{(f)}} 
  \put(20,130) {\small{(g)}}
  \put(160,130) {\small{(h)}}  
  \put(-180,-5) {\small{(i)}}
  \put(-35,-5) {\small{(j)}}
  \put(65,-5) {\small{(k)}}
  \put(160,-5) {\small{(l)}}
\end{picture}
\caption{}
Representative Feynman diagrams for the various perturbative orders
contributing to $4j + \ell\nu$ production at the LHC.
The diagrams in the first row are purely
electroweak, at \ordEW .
In the second row a number of processes at \ordQCD are
illustrated. Diagram (e) refers to one of the main backgrounds, namely $t\bar{t}$
production, while \diags{g}{h} contribute to the $HWjj$ signal at \ordQCD .
The last row exemplifies the $W+4j$ QCD background at \ordQCDsq .
\label{fig:diag}
\end{figure}

Three perturbative orders contribute to $4j + \ell\nu$ at the LHC.
In \fig{fig:diag} some representative Feynman diagrams
are presented. The diagrams in the first row are purely
electroweak: they correspond to boson--boson scattering and to boson--boson
fusion to Higgs with an additional W emission. Contributions similar to 
\diags{f}{g}, with a neutral electroweak boson in place of the gluon, are also
present at \ordEW .
In the second row, \diags{e}{h}, a number of processes at \ordQCD are
illustrated. Diagram (e) refers to one of the main backgrounds, namely $t\bar{t}$
production, while \diags{g}{h} presents some of the possibilities for producing 
a Higgs particle in association with a $W$ boson at \ordQCD .
The last row, \diags{i}{l}, exemplifies the $W+4j$ QCD background at \ordQCDsq
where
no Higgs boson can be present. These processes provide the continuum in the 
mass of the two central jet distribution above which the Higgs signal has to be
extracted.
 
The \ordEW and \ordQCD samples have been generated with \Phantom
\cite{PhantomPaper,method,phact}, while the \ordQCDsq sample 
has been produced with \MadEvent \cite{MadeventPaper}.
Both programs generate events 
in the Les Houches Accord File Format \cite{LHAFF}.
In all samples full $2 \rightarrow 6$ matrix elements, without any
production times decay approximation, have been used.

The \ordQCD contribution is particularly challenging because it is dominated by
$t\overline{t}$ production. It has been necessary to generate two event samples.
The first one has been produced considering only final states with at least two
$b$'s and imposing antitop selection requirements,
in order to obtain a sample in which the
$H\rightarrow b \bar{b}$ peak could be seen. For this purpose,
we have required that no jet triplet satisfies
\be
\label{eq:topcut1}
\vert M_{jjj} - M_t \vert < 15 \,{\mathrm GeV}
\ee
and no jet satisfies 
\be
\label{eq:topcut2}
\vert M_{jl\nu} - M_t \vert < 15 \,{\mathrm GeV}.
\ee
We have taken $M_t = 175 \,{\mathrm GeV}$.
The second sample is the complementary one in which less than two $b$'s
are present in the final state or at least one of the
conditions in \eqns{eq:topcut1}{eq:topcut2} is met.
It includes all contributions
from $t\bar{t}$ and single top production as well as all reactions which cannot
lead to a final state compatible with the production of a Higgs particle
decaying to $b\bar{b}$. These samples will be referred in the
following as $2b-notop$ and $rest$ respectively. The second one is dominated
by top production even though it includes a much larger set of reactions.

The \ordQCDsq sample, which correspond to  $W + 4j$,
includes all possible reactions with $b$'s in the
final state as well as all reactions
without any final $b$, which can only contribute through fake hits.
 
We work at parton level with no showering and hadronization.
The two jets with the largest and smallest rapidity are identified as forward and
backward tag jet respectively. The two intermediate jets are considered as
candidate Higgs decay products.

The neutrino momentum is reconstructed according to the usual prescription,
requiring the 
invariant mass of the $\ell \nu$ pair to be equal to the $W$ boson nominal mass,
\begin{equation}
\label{eq:nu_reco_equation}
(p^{\ell}+p^{\nu})^2 = M_W^2 ,
\end{equation}
in order to determine the longitudinal component of the neutrino momentum.
This equation has two solutions,
\begin{equation}
\label{eq:nu_reco}
p_{z}^{\nu} = \frac{\alpha p_z^\ell \pm \sqrt{\alpha^2 p_z^{\ell 2} - 
 (E^{\ell 2} - p_z^{\ell 2})(E^{\ell 2} p_T^{\nu 2} - \alpha^2)}}
  {E^{\ell 2} - p_z^{\ell 2}}  \; ,
\end{equation}
where
\begin{equation}
\alpha = \frac{M_W^2}{2} + p_x^{\ell}p_x^{\nu} + p_y^{\ell}p_y^{\nu}  \; .
\end{equation}

If the discriminant of Eq.(\ref{eq:nu_reco}) is negative, which happens only if
the actual momenta satisfy $(p^{\ell}+p^{\nu})^2 > M_W^2$,
it is reset to zero.
The corresponding value of $p_{z}^{\nu}$ is adopted.
This value of $p_{z}^{\nu}$ results in the smallest possible value for
the mass of the $\ell \nu$ pair which is compatible with the known componenents
of $p^{\ell}$ and $p^{\nu}$. The corresponding mass is always larger than $M_W$.
If the determinant is positive and the two solutions
for $p_{z}^{\nu}$ have opposite sign we choose the solution whose sign coincides
with that of $p_{z}^{\ell}$. If they have the same sign we choose the
solution with the smallest $\Delta R$ with the charged lepton.
The reconstructed value is used for computing all physical observables.

Our basic selection cuts, which have been applied already in generation are:
\bea
\label{eq:cuts:all}
& p_{T_j} \geq 30~{\rm GeV} \, , \; \; |\eta_j| \leq 5.0 \, , \; \;
 p_{T_\ell} \geq 20~{\rm GeV} \, ,\; \;
|\eta_{\ell}| \leq 3.0 \, ,\nonumber \\
&  M_{jj} \geq 60~{\rm GeV}
\eea
where $j = d,u,s,c,b,g$.
$b$--tagging is active for $|\eta| \leq 2.4$ \cite{CMS-bTag}
with efficiencies $\epsilon_b = 0.5$ for a $b$--quark,
$\epsilon_c = 0.1$ for a $c$--quark, $\epsilon_{q/g} = 0.01$ for a 
light quark or a gluon. It should however be noted that these efficiencies are
up to a point tunable, modifying the identification thresholds, and adapted to
the analysis at hand. This necessarily involves a tradeoff between efficiency
and purity and is expected to improve in parallel with the understanding of the 
detector response.

All samples have been generated using CTEQ5L \cite{CTEQ5} 
parton distribution functions.
For the \ordEW and \ordQCD samples, generated with \Phantom,
the QCD scale has been taken as:
\be
\label{eq:LargeScale}
Q^2 = M_W^2 + \frac{1}{6}\,\sum_{i=1}^6 p_{Ti}^2.
\label{scale}
\ee
while for the \ordQCDsq sample the scale
has been set to $Q^2 = M_Z^2$. This difference in the scales
leads to a definite relative enhancement of
the  $4j\, +  \, W$ background. Tests in
comparable reactions have shown an increase of about a factor of 1.5 
for the processes computed at $Q^2 = M_Z^2$ with
respect to the same processes computed with the larger scale
\eqn{eq:LargeScale}.

In our estimates below we have only taken into account the muon and electron
decays of the $W$ boson. The possibility of detecting high $p_T$ taus has been
extensively studied in connection with the discovery of a light Higgs in Vector
Boson Fusion in the
$\tau^+\tau^-$ channel \cite{ATLAS-HinVV} with extremely encouraging results.
Efficiencies of order 50\% have been obtained for the hadronic decays of the
$\tau's$. The expected number of events in the $H\rightarrow  \tau\tau
\rightarrow e\mu + X$ is within a factor of two of the yield from 
$H\rightarrow  W W^* \rightarrow e\mu + X$ for $M_H = 120$ \GeV where the
$\tau\tau$ and $W W^*$ branching ratios of the Higgs boson are very close,
suggesting that also in the leptonic decay channels of the taus the efficiency
is quite high. Therefore we expect the $W \rightarrow \tau \nu$ channel
to increase the detectability of the $b\bar{b}Wjj $ final state.

A minijet veto has been broadly discussed as a tool to separate electroweak
dominated processes from QCD dominated backgrounds. For the class of processes
we study in  this paper, this issue has been raised in
Refs.\cite{Rainwater00,bbjjgamma}. Both groups foresee large gains in statistical
significance with modest losses in signal rate.

\section{Double b--tagging analysis}
\label{sec:2b}

The two central jets are required to be $b$ tagged within the active region
$|\eta| \leq 2.4$.

As a check,
using cuts similar (but not exactly equal because of the lower $p_T$ threshold
used in \cite{Rainwater00} compared to \eqn{eq:cuts:all} where $b$'s are treated
on the same footing as all other jets) to 
Eq. (\ref{eq:gap}) + Eq. (\ref{eq:mjjptb}) and perfect $b$--jet efficiency 
($\epsilon_b = 1.0$, $\epsilon_c = \epsilon_{q/g} = 0.0$)
we obtain results which are in reasonable agreement with those in the second
column of Table II of \cite{Rainwater00}. In the following we will use a
different selection procedure.

In addition to the basic selection cuts \eqn{eq:cuts:all} the following cuts are
imposed:

\begin{figure}[h]
\begin{center}
\mbox{\includegraphics*[width=12.cm]{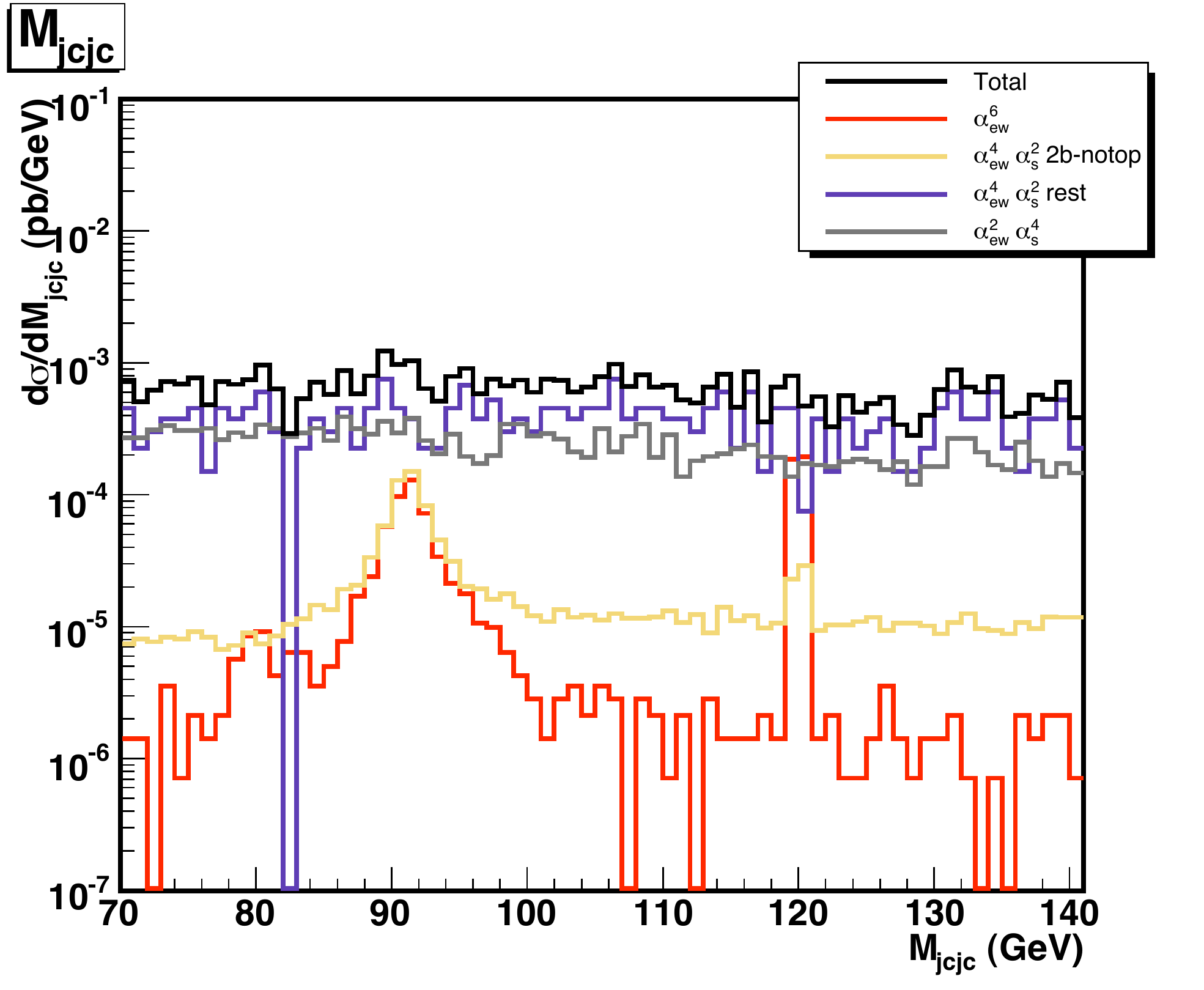}}
\caption {}
Mjj distribution for double b--tagging.
Cuts as in \eqn{eq:cuts:all} and \eqns{eq:cuts:2b1}{eq:cuts:2b4}.
\label{Mjj:2b}
\end{center}
\end{figure}

\be
\label{eq:cuts:2b1}
M_{j_cj_c j_{1(2)}} \geq 185~{\rm GeV} 
\ee
\be
\label{eq:cuts:2b2}
\triangle \eta_{tags} = |\eta_{j_1}-\eta_{j_2}| \geq 1.8 
\ee
\be
\label{eq:cuts:2b3}
M_{j_1 j_2} \geq 400~{\rm GeV}  
\ee
\be
\label{eq:cuts:2b4}
 \triangle \eta_{_{VV}} \geq 0.7 
\ee
where $j_cj_c$ refers to the two central jets and $\triangle \eta_{_{VV}}$
is the pseudorapidity separation between the reconstructed vector bosons.

\begin{table*}[h]
\label{Xsection:2b}
\vspace{0.15in}
\begin{center}
\begin{tabular}{|c|c|c|c|c|c|}
\hline
 $W+4j$  & $\;$ $\;$(\ref{eq:cuts:all})$\;$ $\;$ &  + (\ref{eq:cuts:2b1}) 
  &  + (\ref{eq:cuts:2b2})  & + (\ref{eq:cuts:2b3})  & + (\ref{eq:cuts:2b4})   \\
\hline
\ordEW signal   & 90 & 64 & 59 & 50 & 41  \\
\hline
\ordQCD$_{2b-notop}$  & 54 & 53 & 48 & 37 & 25  \\
\hline
\ordQCD$_{rest}$  & 9168 & 3332 & 1606 & 1108 & 641  \\
\hline
\ordQCDsq   & 770 & 705 & 625 & 425 & 365 \\
\hline
$S/\sqrt{B}$ & 0.90 & 1.00 & 1.24  & 1.26 & 1.28 \\
\hline
\end{tabular}
\end{center}
\caption {}
Number of events for an integrated luminosity L = 100 fb$^{-1}$ for
each leptonic decay channel of the $W$ for the $WHjj$ signal with
$M_H = 120$~GeV, and the principal backgrounds.
The two central jets are required to be $b$--tagged.
The selection cuts are given in 
\eqn{eq:cuts:all} and \eqns{eq:cuts:2b1}{eq:cuts:2b4}. 
A mass bin $110 < M_{b\bar{b}} < 130$~GeV is assumed.
The statistical significancies are obtained considering only the \ordEW events
in the $M_H \pm 10$~GeV mass window
as the signal and everything else as background.
\end{table*}

The initial sample of \ordEW events 
contains a non negligible contribution from $t\bar{t}$
purely EW production which is eliminated by the requirement 
$M_{j_cj_c j_{1(2)}} \geq 185~{\rm GeV}$.
 
More stringent cuts on 
$|\eta_{j_1}-\eta_{j_2}|$ or on $M_{j_1 j_2}$,
as employed for instance in $WW$ scattering studies,
do not improve the statistical significance.
\fig{Mjj:2b} shows that the $Wb\bar{b}jj$ \ordQCD$_{2b-notop}$ contribution
to the
signal turns out to be small, while adding a few events to the background count.

The \ordEW cross section is dominated by the Higgs and $Z$ peaks, with a small
background contribution due to misidentification of the Higgs decay products,
jet mistagging and the irreducible background from diagrams which do not include
Higgs production as a subdiagram.
It is therefore
appropriate to consider the integral of the mass distribution in the
$\pm 10 \; \GeV$ mass window around the Higgs mass as the signal, as we do in
Table 1 
and in the following.
The \ordEW background discussed above amounts to about 10\% in the selected mass window.

The total
background is essentially flat in the mass region of interest. It can be
measured from the sidebands of the known Higgs mass, drastically decreasing
the measurement uncertainty. 
Even though we have not explicitely required a minimum $\Delta R$ 
among jets, they turn out to be well separated. Only about 5\% of the events
which pass all selection cuts have
$\Delta R < 0.5$ in each sample.

The contribution of the $t\bar{t}jj$ background is estimated in
\cite{Rainwater00} to be at most of a few hundredths of a femtobarn per GeV.
Therefore, being much smaller than the  $t\bar{t}$ one, has been neglected.  

Considering only the \ordEW events in the $M_H \pm 10 GeV$ mass window
as the signal, the significance $S/\sqrt{B}$ is about 1.28.
A factor of two can be gained by considering the $W$ decay to
both $e$ and $\mu$ and summing the statistics of the
two experiments.  
Obtaining the standard 5$\sigma$ significance requires a further factor of about
3.8 in statistics.

Our estimate is considerably more pessimistic than the one presented in
\cite{Rainwater00}, with a signal over background ratio of about 1/25 instead of
about 1/7, with a similar number of expected signal events in the two analysis.
This is due to to the inclusion of the $t\bar{t}$ channel which represent
about two thirds of the background and also to a larger predicted background
from $Wjjjj$  due to the effect of $c$-quarks and light partons being tagged as
$b$'s.

This unsatisfactory result depends quite strongly on the assumed tagging and
fake probabilities. Any improvement in this area could make this channel much
more attractive.

Since the background is dominated by QCD processes a minijet veto could be
useful.

A more sophisticated selection based on a multivariate data analysis is likely
to yield more optimistic results.

The \ordEW cross section for $Z(\rightarrow \!b\bar{b})Wjj$ is of the same order of
magnitude as the cross section for $H(\rightarrow \!b\bar{b})Wjj$.
Therefore it might be
possible to observe $ZWjj$ production through the $Z$ decay to $b\bar{b}$.
This can be exploited as a calibration/control point and as a
window to ZW scattering which is difficult to separate from WW scattering in the
semileptonic channel.

We can compare our result with those presented in Table 8 of
Ref.\cite{bbjjgamma} 
where both irreducible and reducible backgrounds are taken into account.
In \cite{bbjjgamma} the $b$--tagging efficiency is taken as $\epsilon_b = 0.6$,
while a reduction of the signal events by 70\% is assumed as a consequence of
finite $b\bar{b}$ mass resolution. The combined effect of a larger $\epsilon_b$
and of including the $b\bar{b}$ mass resolution nicely corresponds to taking
$\epsilon_b = 0.5$ and the two set of results are directly comparable.
The expected significance in the $b\bar{b}\gamma jj$ channel is 2.2(1.8) for
$p_T^\gamma > 20(30)$ \GeV which is to be compared with a significance of 1.81
in the $b\bar{b}\ell\nu jj$ channel, when two lepton species
$\ell = e,\; \mu$ are considered. The number of
expected signal events is similar in all cases.

\section{Single b--tagging analysis}
\label{sec:1b}

At least one of the two central jets is required to be  $b$--tagged within the
active region $|\eta| \leq 2.4$. The second central jet can fall anywhere in the
region $|\eta| \leq 5.0$.

\begin{table*}[bth]
\label{Xsection:1b}
\vspace{0.15in}
\begin{center}
\begin{tabular}{|c|c|c|c|c|c|}
\hline
 $W+4j$  & $\;$ $\;$(\ref{eq:cuts:all})$\;$  $\;$ &  + (\ref{eq:cuts:1b1}) 
  &  + (\ref{eq:cuts:1b2})  & + (\ref{eq:cuts:1b3})  & + (\ref{eq:cuts:1b4})\\
\hline
\ordEW signal       & 1037 & 285 & 146 & 122 & 99  \\
\hline
\ordQCD$_{2b-notop}$ & 335 & 322 & 126 & 96 & 69  \\
\hline
\ordQCD$_{rest}$ & 177490 & 24036 & 4328 & 3611 &  2352  \\
\hline
\ordQCDsq        & 19584 & 17408 & 7354 & 4636 & 3487 \\
\hline
$S/\sqrt{B}$ & 2.33 & 1.39 & 1.34 & 1.34 & 1.29\\
\hline
\end{tabular}
\end{center}
\caption {}
Number of events for an integrated luminosity L = 100 fb$^{-1}$ for
each leptonic decay channel of the $W$ for the $WHjj$ signal with
$M_H = 120$~GeV, and the principal backgrounds.
At least one of the two central jets is required to be $b$--tagged.
The selection cuts are given in 
\eqn{eq:cuts:all} and \eqns{eq:cuts:1b1}{eq:cuts:1b4}. 
A mass bin $110 < M_{b\bar{b}} < 130$~GeV is assumed.
The statistical significancies are obtained considering only the \ordEW events
in the $M_H \pm 10 $~GeV mass window as the signal and everything else as
background.
\end{table*}

In addition to the basic selection cuts \eqn{eq:cuts:all} the following cuts are
imposed:
\be
\label{eq:cuts:1b1}
M_{j_cj_c j_{1(2)}} \geq 185~{\rm GeV}
\ee
\be
\label{eq:cuts:1b2}
M_{j_1 j_2} \geq 600~{\rm GeV} 
\ee
\be
\label{eq:cuts:1b3}
\triangle \eta_{tags} = |\eta_{j_1}-\eta_{j_2}| \geq 4.0 
\ee
\be
\label{eq:cuts:1b4}
M_{vis} \geq 1200~{\rm GeV} 
\ee
Here $M_{vis}$ is the total visible mass, which at parton level coincides with
the mass of the $4j + \ell$ system.
 
\begin{figure}[ht]
\begin{center}
\mbox{\includegraphics*[width=12.cm]{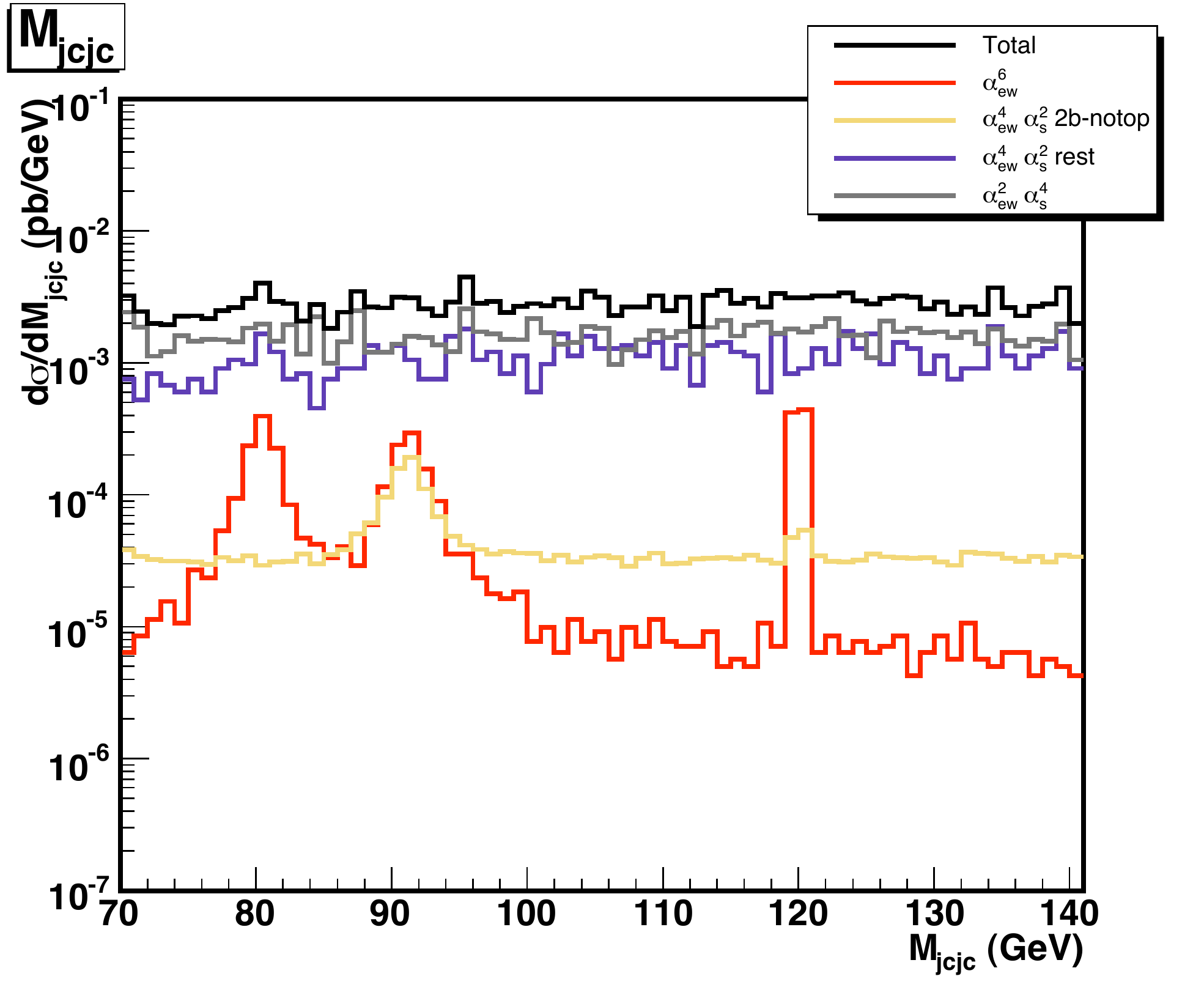}}
\caption {}
Mjj distribution for single b--tagging.
Cuts as in \eqn{eq:cuts:all} and \eqns{eq:cuts:1b1}{eq:cuts:1b4}.
\label{Mjj:1b}
\end{center}
\end{figure}

Both a $Z$ and a $W$ peak appear in the invariant mass distribution in
\fig{Mjj:1b}. Their cross section is of the same order of magnitude as the Higgs
one and might therefore be observable.

Also in this case, the background is essentially flat and a direct 
measurement from the sidebands is possible. 
As mentioned before, the $Wjjjj$  background, which is the largest one in the
present case, has been generated with a smaller
QCD scale compared to the other samples. Therefore our estimate of this
background is quite conservative.
In the following we will only take into account events in a
$\pm 10 \; \GeV$ mass window around the Higgs mass.
Considering only the \ordEW events as the signal, the significance $S/\sqrt{B}$
is about 1.29 
which becomes 1.58 if we assume that the $W + 4j$ background is
overestimated to about 1.5 times its actual value.
A factor of two can be gained by considering the $W$ decay to both
$e$ and $\mu$ and summing the statistics of the
two experiments.  
Obtaining the standard 5$\sigma$ significance requires a further factor of about
3.8 in statistics.

The significancies reported in the last line of 
Table 2, 
and in particular their decrease as the cuts are tightened, should be
taken with great care. In fact,
there is in general a significant contamination of what
we define as signal from the \ordEW continuum which decreases as further cuts
are imposed. As can be readily seen from \fig{Mjj:1b} this contamination is
about 15\% for the rightmost column. For the first column on the left the
continuum contribution is actually about three times the area of the Higgs peak.
Therefore while the significancies are coherent with our definition of signal,
they do not actually correspond, at least for the columns on the left side,
to the ratio of the pure Higgs signal to the square root of the total
background.

To our knowledge, a single $b$--tagging approach to the detection of the Higgs
decay to $b\bar{b}$ has only been briefly discussed in \cite{bbjjgamma} but no
detailed analysis has been reported. Our study suggests that such an approach is
perfectly viable even when all backgrounds are included and that indeed it results
in the same statistical
significance as the selection procedure that requires two jets to be
tagged. We expect this technique to be
useful in other production channels as the $b\bar{b}\gamma jj$ one.

Again a minijet veto could be useful. 

It is obviously feasible to combine the double tag and single tag measurements.
It is possible,
requiring for instance exactly one tag in the single tag analysis, to produce 
event samples which are mutually exclusive. Nonetheless they would be
correlated because of common systematic uncertainties, as a consequence, for
example, of using the same procedure for tagging $b$'s.
Therefore it is non trivial to give an estimate of the statistical significance
of the combined measurements in the double tag and single tag channels. 

\section{No b--tagging analysis}
\label{sec:0b}

As in the previous analysis the background due to the top appears to be
manageable. However in this case the \ordQCDsq contribution, $W+4j$, is
overwhelming. We have been unable to find physical observables which provide
sufficient discrimination. This channel therefore looks hopeless.

\begin{table*}[h]
\label{Xsection:all}
\vspace{0.15in}
\begin{center}
\begin{tabular}{|c|c|c|}
\hline
 $W+b\bar{b}+2j$  & 2 tags & 1 tag   \\
\hline
signal   &  41  & 99 \\
\hline
background  & 1031 & 5908\\
\hline
$S/\sqrt{B}$  & 1.28 & 1.29 \\
\hline
$S/\sqrt{B}$  & 3.13 & 3.15 \\
\hline
\end{tabular}
\end{center}
\caption {}
Number of events and statistical significancies for the $WHjj$ signal with
$M_H = 120$~GeV and the total background in the two tag and single tag
approach.
The first three lines refer to an integrated luminosity L = 100 fb$^{-1}$ for
each leptonic decay channel of the $W$.
The fourth line gives the corresponding statistical significancies for
an integrated
luminosity L = 300 fb$^{-1}$ and summing over both leptonic channels.
In addition to the basic selection cuts \eqn{eq:cuts:all} the cuts in 
\eqns{eq:cuts:2b1}{eq:cuts:2b4} and \eqns{eq:cuts:1b1}{eq:cuts:1b4}
respectively are applied to the two tag and single tag case.
A mass bin $110 < M_{b\bar{b}} < 130$~GeV is assumed.
The statistical significancies are obtained considering only the \ordEW events
in the $M_H \pm 10 $~GeV mass window
as the signal and everything else as background.
\end{table*}

\section{$WHjj$ production at the SLHC}
\label{sec:SHLC}
Since the number of expected events for the projected LHC total luminosity of
about 300 fb$^{-1}$ is rather small, the reaction we have discussed in this
paper could benefit from the larger luminosity which would be provided by the 
Super LHC which is expected to be ten times higher.  
According to the preliminary studies reported in \cite{SLHC-rep}, it can be
assumed that at the SLHC $b$--tagging and reconstruction of isolated high
$p_T$ particles will be possible with efficiencies comparable to those
presently foreseen
at CMS and ATLAS. The most serious challenge for all vector boson fusion like
channels, which rely on tag jets in order to discriminate between signal and
QCD dominated background, is represented by pile--up: multiple scattering events
during the same bunch crossing which could produce spurious jets unrelated
to the main interaction.
These jets could mimic forward and backward tag
jets as well as spoil the possibility of a central jet veto to increase the
signal
to background ratio. The probability of extra jets from pile--up depends
strongly
on the energy or $p_T$ threshold, therefore it is generally expected that a good
efficiency can be recovered using more stringent requirement on jet recognition.
Quite interestingly, another route to the same goal is to adopt a smaller cone
size. This approach would benefit the whole field of vector boson fusion
processes in which high $p_T$ bosons decay to two jets which often merge into
one when cones of size $\Delta R = 0.5 \div 0.7$ are used.

\begin{figure}[ht]
\begin{center}
\mbox{\includegraphics*[width=12.cm]{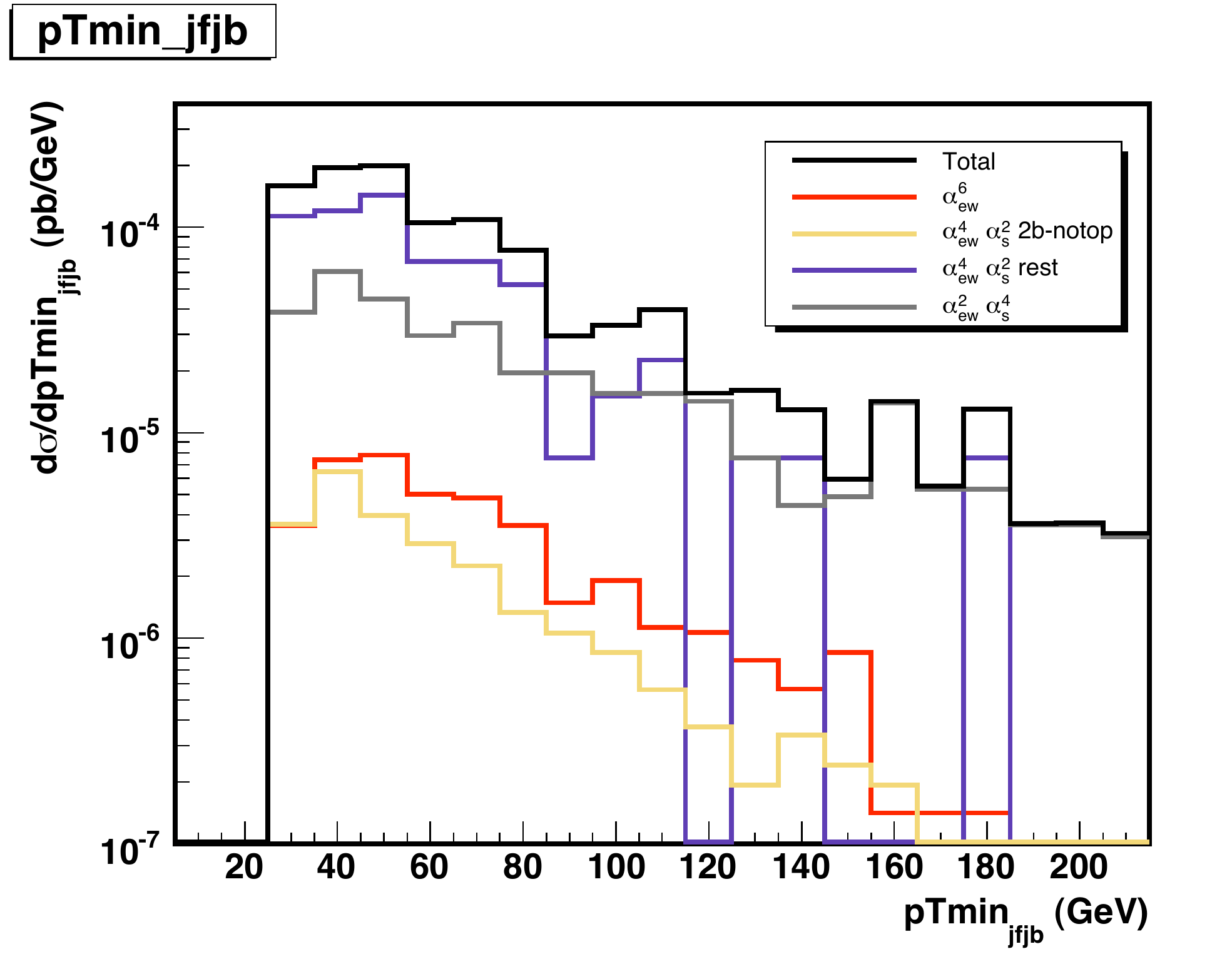}}
\caption {}
Distribution of the smallest $p_T$ of the two tag jets in the double b--tagging
case.
Cuts as in \eqn{eq:cuts:all} and \eqns{eq:cuts:2b1}{eq:cuts:2b4}.
\label{pTmin-jfjb:2b}
\end{center}
\end{figure}

While a detailed study of pile--up effects is beyond the scope of this paper,
in \fig{pTmin-jfjb:2b} we present the distribution of the smallest $p_T$ of
the two tag jets in the double b--tagging case. The distribution is similar
for all set of processes, with the maximum at about 50 \GeV and then a rather
mild decrease at larger values. For instance, setting the threshold at 100 \GeV
would lead to 120 signal event and about 3500 events for the background,
for an integrated
luminosity L = 3000 fb$^{-1}$ and summing over both leptonic channels, with a
significance of about two. Clearly, the $p_T$ threshold and the full selection
procedure can be optimized however, taking into account that a large additional
background from pile--up is expected, the prospect of studying $WHjj$ production
at the SLHC looks quite difficult.     

\section{Conclusions}
\label{sec:conclusions}

In this paper we have studied Higgs production in association with a $W$ boson
and two energetic tag jets at the LHC for $M_H = 120$ \GeV, with the Higgs decaying to
$b\bar{b}$ and the $W$ to leptons.
All parton level backgrounds have been analyzed, including the effect of fake
$b$--tagging which, with realistic efficiencies, turns out to be large.
We have discussed two detection strategies, whose results are summarized in
Table 3: 
the first one requires two jets to
be $b$--tagged. In the second one, which has not been examined in detail before, 
at least one tag is required. After all selection cuts about 80 and 200
events are predicted in the two cases, summing the electron and muon channels,
for a standard luminosity of 100 $fb^{-1}$
with a S/B ratio of 1/25 and 1/60 respectively. 
The corresponding statistical significancy, S/$\sqrt{\mathrm B}$, 
are of the order of 1.3 for each leptonic decay channel of the $W$ boson with 
a luminosity of 100 $fb^{-1}$, which becomes of order three for
L = 300 fb$^{-1}$ and summing over both leptonic channels,
comparable to the significancies found in other channels.
For the double $b$--tagging case
looser cuts than those used in previous studies appear to be viable.
The single $b$--tagging analysis provides the same 
statistical significance than the double $b$--tagging one  
with almost three times the number of signal events.
It appears unlikely that $WHjj$ production could be studied at the SLHC 
exploiting the higher luminosity.

\section *{Acknowledgments}
A.B. wishes to thank the Department of Theoretical Physics of Torino University for support.

This work has been supported by MIUR under contract 2006020509\_004 and by the
European Community's Marie-Curie Research 
Training Network under contract MRTN-CT-2006-035505 `Tools and Precision
Calculations for Physics Discoveries at Colliders'

\newpage


\end{document}